\documentclass[pdflatex,sn-aps]{sn-jnl}

\usepackage{amsmath}
\usepackage{siunitx, physics}
\usepackage{eurosym}
\usepackage{bm,upgreek}
\usepackage{url}
\jyear{2022}%

\theoremstyle{thmstyleone}%
\theoremstyle{thmstyletwo}%

\theoremstyle{thmstylethree}%

\begin{document}

\title[Portable MRI of patients indoors, outdoors and at home]{Portable magnetic resonance imaging of patients indoors, outdoors and at home}


\author[1]{\fnm{Teresa} \sur{Guallart-Naval}}
\equalcont{These authors contributed equally to this work.}
\author[2,3]{\fnm{Jos\'e~M.} \sur{Algar\'in}}
\equalcont{These authors contributed equally to this work.}
\author[4,5]{\fnm{Rub\'en} \sur{Pellicer-Guridi}}
\author[2,3]{\fnm{Fernando} \sur{Galve}}
\author[4,6]{\fnm{Yolanda} \sur{Vives-Gilabert}}
\author[1]{\fnm{Rub\'en} \sur{Bosch}}
\author[2,3]{\fnm{Eduardo} \sur{Pall\'as}}
\author[1]{\fnm{Jos\'e~M.} \sur{Gonz\'alez}}
\author[1]{\fnm{Juan~P.} \sur{Rigla}}
\author[1]{\fnm{Pablo} \sur{Mart\'inez}}
\author[4]{\fnm{Francisco~J.} \sur{Lloris}}
\author[1]{\fnm{Jose} \sur{Borreguero}}
\author[7]{\fnm{\'Alvaro} \sur{Marcos-Perucho}}
\author[8]{\fnm{Vlad} \sur{Negnevitsky}}
\author[9]{\fnm{Luis} \sur{Mart\'i-Bonmat\'i}}
\author[1]{\fnm{Alfonso} \sur{R\'ios}}
\author[2,3]{\fnm{Jos\'e~M.} \sur{Benlloch}}
\author*[2,3]{\fnm{Joseba} \sur{Alonso}}\email{joseba.alonso@i3m.upv.es}

\affil[1]{\orgname{Tesoro Imaging S.L.}, \orgaddress{\city{Valencia}, \postcode{46022}, \country{Spain}}}
\affil[2]{\orgdiv{Institute for Molecular Imaging and Instrumentation}, \orgname{Spanish National Research Council}, \orgaddress{\city{Valencia}, \postcode{46022}, \country{Spain}}}
\affil[3]{\orgdiv{Institute for Molecular Imaging and Instrumentation}, \orgname{Universitat Polit\`ecnica de Val\`encia}, \orgaddress{\city{Valencia}, \postcode{46022}, \country{Spain}}}
\affil[4]{\orgname{PhysioMRI Tech S.L.}, \orgaddress{\city{Valencia}, \postcode{46022}, \country{Spain}}}
\affil[5]{\orgname{Asociaci\'on de investigaci\'on MPC}, \orgaddress{\city{San Sebasti\'an}, \postcode{20018}, \country{Spain}}}
\affil[6]{\orgdiv{Department of Electronic Engineering}, \orgname{Universitat de Val\`encia}, \orgaddress{\city{Burjassot}, \postcode{46100}, \country{Spain}}}
\affil[7]{\orgname{Helios School}, \orgaddress{\city{La Eliana}, \postcode{46183}, \country{Spain}}}
\affil[8]{\orgname{Oxford Ionics}, \orgaddress{\city{Oxford}, \postcode{OX5 1PF}, \country{United Kingdom}}}
\affil[9]{\orgdiv{Medical Imaging Department}, \orgname{Hospital Universitari i Polit\`ecnic La Fe}, \orgaddress{\city{Valencia}, \postcode{46026}, \country{Spain}}}

\abstract{Mobile medical imaging devices are invaluable for clinical diagnostic purposes both in and outside healthcare institutions\cite{Nelson2011,Pittayapat2010,Kosaka2009,Zubair2021}. Among the various imaging modalities, only a few are readily portable. Magnetic resonance imaging (MRI), the gold standard for numerous healthcare conditions\cite{Smith2010,Gordillo2013,Jack2008,Carlsson2008,Anthoulakis2014,Pawlyn2015}, does not traditionally belong to this group. Recently, low-field MRI start-up companies have demonstrated the first decisive steps towards portability within medical facilities\cite{Sheth2021,Mazurek2021,Nasri2021}, but these are so far incompatible with more demanding use cases such as in remote and developing regions, sports facilities and events, medical and military camps, or home healthcare. Here we present \emph{in vivo} images taken with a light, home-made, low-field extremity MRI scanner outside the controlled environment provided by medical facilities. To demonstrate the true portability of the system and benchmark its performance in various relevant scenarios, we have acquired images of a volunteer's knee in: i) an MRI physics laboratory; ii) an office room; iii) outside a campus building, connected to a nearby power outlet; iv) in open air, powered from a small fuel-based generator; and v) at the volunteer's home. All images have been acquired within clinically viable times, and signal-to-noise ratios (SNR) and tissue contrast suffice for 2D and 3D reconstructions with diagnostic value, with comparable overall image quality across all five situations. Furthermore, the volunteer carries a fixation metallic implant screwed to the femur, which leads to strong artifacts in standard clinical systems\cite{Ludeke1985,Harris2006,Stradiotti2009} but appears sharp in our low-field acquisitions. Altogether, this work opens a path towards highly accessible MRI under circumstances previously unrealistic.}


\maketitle

The average adult human body contains over four thousand trillion trillion hydrogen nuclei, each a minuscule quantum compass needle susceptive to the presence of external magnetic fields. Standard clinical MRI scanners make use of powerful superconducting magnets that interact strongly with these protons and enable the high SNR and image contrast and resolution typical for magnetic resonance images\cite{BkHaacke}. Regrettably, these magnets also require cryogenic refrigeration, they are bulky, heavy, environmentally unfriendly, expensive to build, site, operate and maintain, and they ultimately constitute a formidable barrier to the accessibility and democratization of MRI\cite{Marques2019,Sarracanie2020,Wald2020}. Besides, high-field scanners are constrained by patient safety considerations, due to increased specific absorption rates (SAR) of electromagnetic energy in tissues at the corresponding higher excitation radio-frequencies (RF)\cite{Panych2018}; they generate undesirable acoustic noise due to strong magnetic interactions during imaging sequences\cite{Price2001}; and they induce severe image artifacts around metallic implants due to magnetic susceptibility effects\cite{Ludeke1985,Harris2006,Stradiotti2009}. Low-field systems ($<0.3$\,T) can overcome all of the above and are nowadays gaining traction as affordable complements to standard MRI scanners. Recent achievements with low-field scanners include \emph{in vivo} brain and extremity imaging\cite{Cooley2020,OReilly2020}, hard-tissue imaging\cite{Algarin2020,Gonzalez2021,Borreguero2022} and even quantitative MRI and fingerprinting\cite{OReilly2021,Sarracanie2021}. The main penalty to pay for operating in this regime is a significant loss in SNR and spatial resolution. However, the diagnostic value of the resulting reconstructions is not necessarily compromised, due to a number of reasons: i) contrast-to-noise ratio (CNR), a more relevant metric for diagnosis than SNR, does not depend as strongly on field strength\cite{Ghazinoor2007}; ii) multiple health conditions and diseases may be diagnosed without the exquisite detail provided by high-field images\cite{Marques2019}; iii) SAR effects can be safely disregarded at low fields, allowing for efficient pulse sequences which increase the duty cycle to partly compensate the SNR loss\cite{Marques2019}; and iv) machine learning and deep convolutional neural networks can be trained to reconstruct high quality images from low-field acquisitions by e.g. transfer learning\cite{Koonjoo2021,GarciaHernandez2021}.

The scope of conceivable applications for MRI technologies widens extraordinarily once the need for large superconducting magnets is removed. For instance, a vehicle has been equipped with a 0.2\,T system large enough for a human arm\cite{Nakagomi2019}, and point-of-care and bedside neuroimaging have been demonstrated with a 64\,mT FDA-cleared scanner portable within a clinical facility\cite{Sheth2021,Mazurek2021}. Low-cost devices with improved mobility would open the door to expanding MRI applications to home and hospice care, small clinics, rural areas or sports clubs and school facilities. Autonomously powered scanners could even be operated outdoors, e.g. in sports events, campaign hospitals, NGO and military camps\cite{Sarracanie2015} or medicalized vehicles\cite{Nakagomi2019}, making MRI available to a large fraction of the world population with no or insufficient access\cite{Marques2019,Sarracanie2020,Wald2020}. 

In this article we present a 70\,mT extremity MRI scanner mounted on a wheeled structure, with an overall weight $\approx250$\,kg and component cost $<50$\,k\euro. After checking that the system performs as expected for \emph{in vivo} images under controlled ambient conditions in an MRI physics laboratory, we took images of the right knee of a volunteer in different indoor and outdoor environments, including the living room at the volunteer's apartment, and in open air connected to a portable gasoline generator. All knee images were acquired with identical 3-dimensional turbo spin echo (3D-TSE) sequences, in about 12\,min each. The electromagnetic interference (EMI) spectrum was different at the various locations, which results in slightly different noise patterns in the reconstructed images. Nevertheless, they all yield valuable anatomical information in clinically acceptable times. The volunteer had undergone a femoral shaft osteotomy and carried a fixation metallic implant screwed to the femur. This hardware is sharply defined in our low-field acquisitions, where previous high-field images suffered from strong susceptibility-induced image distortions.

\begin{figure}
	\centering
	\includegraphics[width=1.\columnwidth]{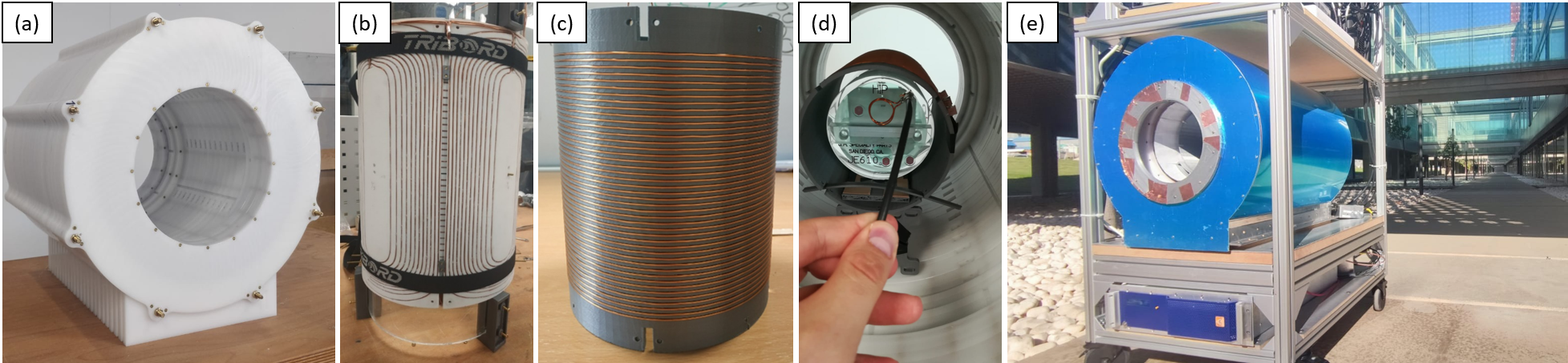}
	\caption{Photographs of the low-field extremity scanner: (a) 70\,mT Halbach magnet; (b) gradient assembly; (c) RF Tx/Rx coil; (d) view of the scanner inside with a phantom in place; and (e) full system mounted on a transportable structure and in open air.}
	\label{fig:ScannerParts}
\end{figure}

The system (Fig.\,\ref{fig:ScannerParts}) is built around a permanent magnet array in a Halbach configuration \cite{OReilly2020}, for a field strength of $\approx70$\,mT homogeneous down to $\approx3,000$\,ppm (parts-per-million) over a diameter of spherical volume (DSV) of 20\,cm. The gradient coil assembly can generate roughly linear fields of up to 40\,mT/m in the DSV. All the below images are taken with a single RF coil used for both transmission and reception (Tx/Rx), tuned and impedance-matched to $\approx3$\,MHz, the proton resonance (Larmor) frequency at our field strength. The control electronics are based on MaRCoS, an open-source, high-performance Magnetic Resonance Control System\cite{Negnevitsky2021,Craven-Brightman2021}. The diameter and length of the scanner are around 53 and 51\,cm respectively, excluding electronics and the mobile structure, with a bore opening $\approx20$\,cm and a weight of $\approx200$\,kg. Once on the mobile, open structure and equipped with all the required electronics and the control computer, the overall system dimensions are $70\times 88\times 166$\,cm$^3$, the weight is $\approx 250$\,kg, and the power consumption is $\ll 1$\,kW. Further details on the scanner can be found in the Methods.

The scanner is usually in the controlled environment of an MRI physics laboratory, where the temperature is stabilized at $18.0\pm0.2$\,C and the air relative humidity at $45\pm10$\,\%. In these conditions, the Larmor frequency is stable down to the kilo-hertz at 3.076\,MHz over weeks, and the different couplings of the subjects to the RF circuitry can be compensated for with minor corrections to the impedance matching electronics. The abundant surrounding electronic equipment and scanners generate substantial EMI in the laboratory at frequencies within our detection bandwidth. For this reason, we conceal the resonant RF coil behind three grounded shields: one is the outermost scanner housing (blue in Fig.\,\ref{fig:ScannerParts}(e)); another is inside the magnet, between the RF coil and the gradient assembly; and last is an electrically conducting cloth that can be wrapped around the subject at both scanner ends, to avoid antenna effects that otherwise couple EMI to the coil from the inside, despite the other two shields.

\begin{figure}
	\centering
	\includegraphics[width=1.\columnwidth]{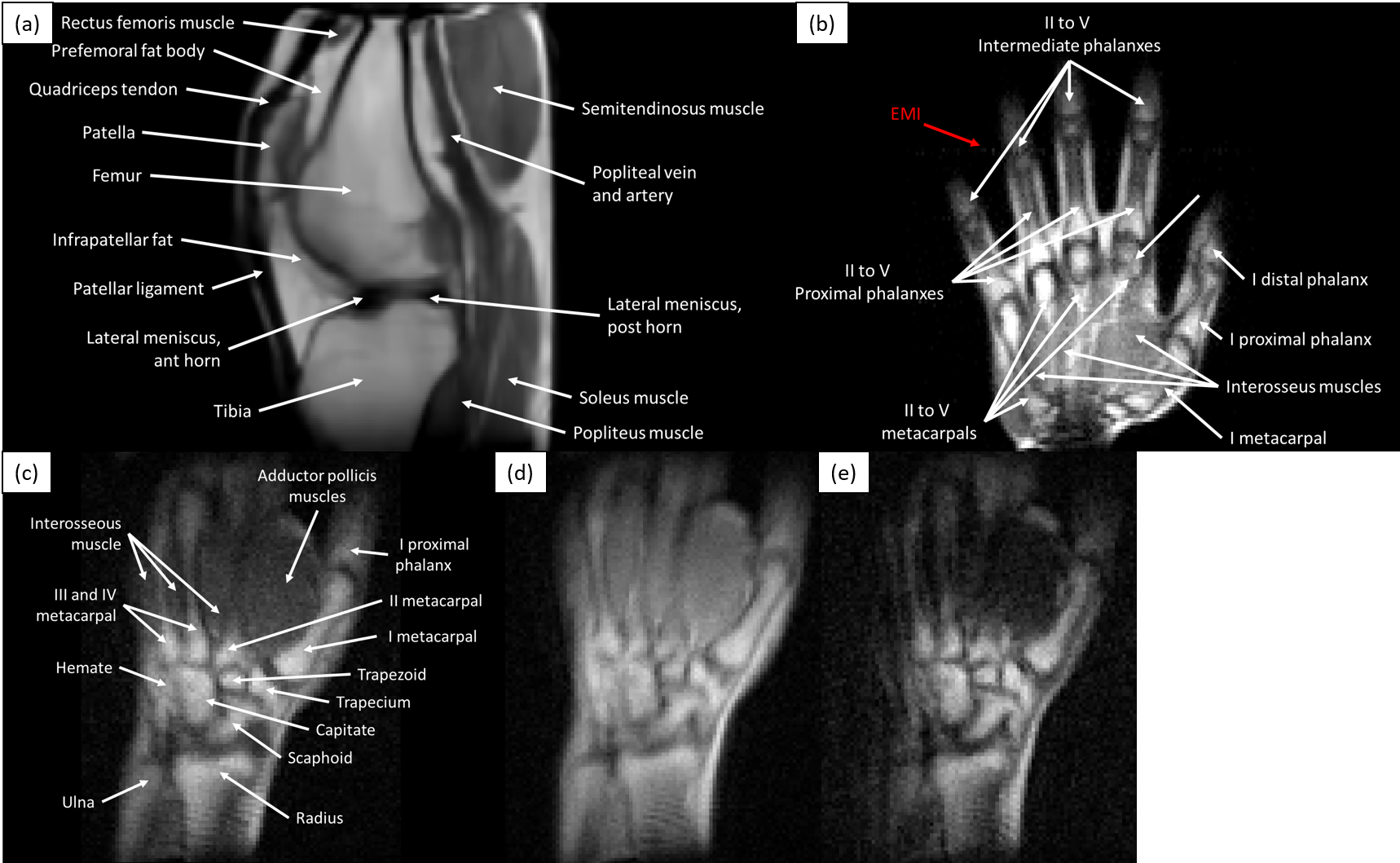}
	\caption{Single slices of 3D-TSE \emph{in vivo} acquisitions of different volunteers in the MRI physics laboratory: (a) $T_1$-weighted image of a knee, acquired in 19\,min, with BM4D filtering and $\times 2$ rescaling; (b) $T_1$-weighted image of a hand (10\,min), without post-processing, with a faint EMI line visible along the phase-encoded direction; (c)-(e) $T_1$, $\rho$ and $T_2$-weighted images of a wrist (12\,min).}
	\label{fig:KneeHand}
\end{figure}

The images in Fig.\,\ref{fig:KneeHand} show the scanner performance in the laboratory. They correspond to \emph{in vivo} 3D-TSE acquisitions of different healthy subjects on different days, showing selected slices of a left knee, a right hand and a right wrist. Acquisition times ranged from 10 to 19\,min (Methods). All images show sufficient tissue contrast and spatial resolution to identify relevant anatomical features, including muscles, fat, cortical bone, bone marrow, tendons, ligaments, veins, arteries and fascia. In these images we show different contrast mechanisms\cite{BkHaacke}, with weightings on $T_1$, $T_2$ and proton density ($\rho$). The knee image in Fig.\,\ref{fig:KneeHand}(a) is BM4D-filtered\cite{Maggioni2013} and rescaled to increase the number of pixels by a factor of $\times 2$, whereas the hand in Fig.\,\ref{fig:KneeHand}(b) is unprocessed after Fourier reconstruction and weak EMI effects result in a faint line along the horizontal (phase-encoded) direction.

\begin{figure}
	\centering
	\includegraphics[width=0.8\columnwidth]{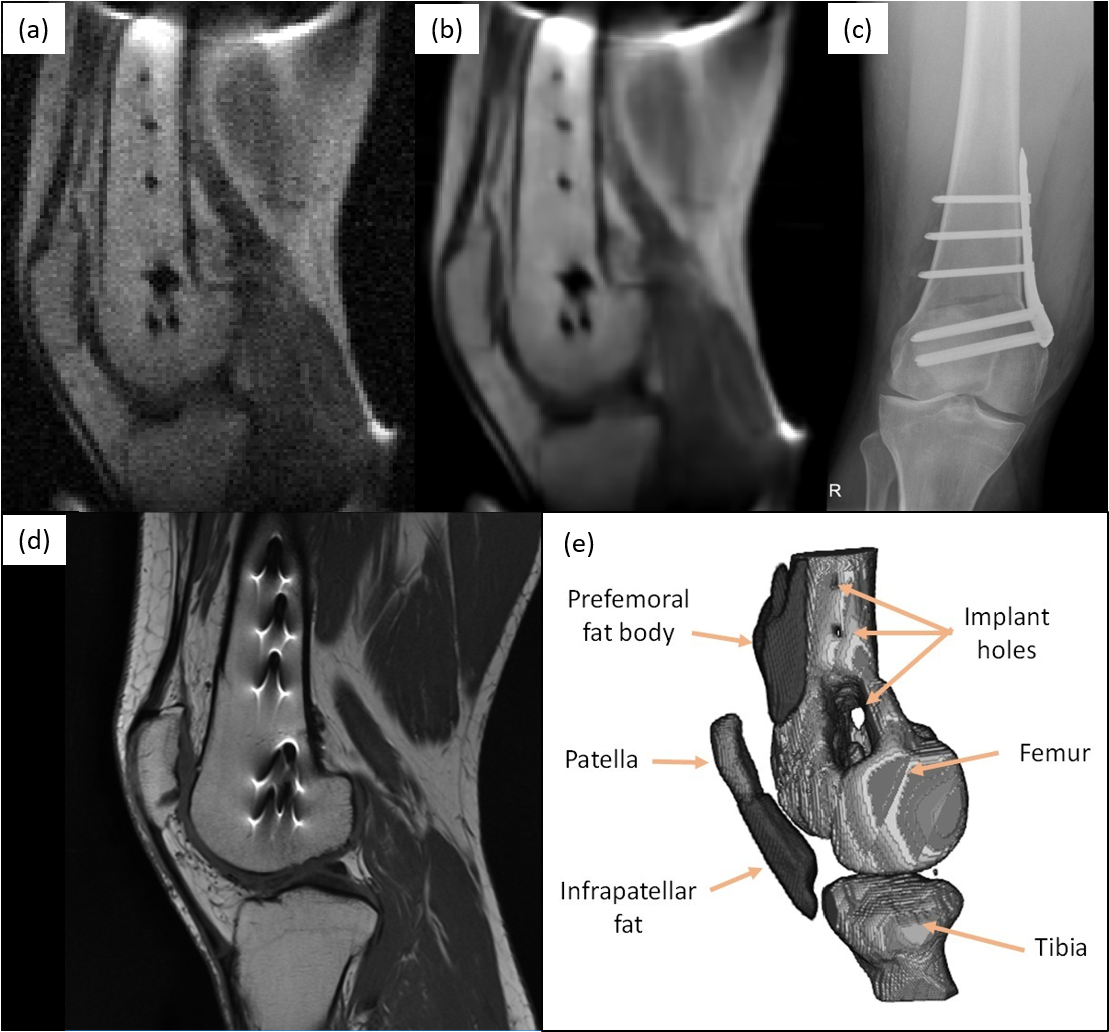}
	\caption{Images of fixation metallic implant attached to the femur, consisting of a plate and seven screws: (a) sagittal view of a raw low-field image acquired with the 70\,mT system (single slice from $T_1$-weighted 3D-TSE acquisition with in-plane resolution of $1.3\times2$\,mm$^2$, 12\,min scan time, eight years after femoral shaft osteotomy); (b) same, but BM4D-filtered and rescaled by $\times 2$; (c) lateral X-ray computed radiography (two weeks after surgery); (d) sagittal view of the same knee, acquired with a Siemens Skyra 3\,T system ($T_1$-weighted 2D-TSE acquisition with slice thickness 3.9\,mm and pixel resolution $0.26\times0.26$\,mm$^2$, one year after surgery); and (e) 3D reconstruction from $T_1$-weighted 3D-TSE acquisition with isotropic resolution of 2\,mm, 20\,min scan time, where selected muscle and fat segments have been removed (eight years after surgery).}
	\label{fig:Implants}
\end{figure}

In a second set of experiments, we demonstrate \emph{in vivo} MR images in the presence of metallic implants without the strong susceptibility-induced artifacts typical of high-field acquisitions\cite{Ludeke1985,Stradiotti2009}, which often hamper post-operative assessment of orthopedic procedures\cite{Harris2006}. The volunteer for these tests had been diagnosed with lateral gonarthrosis due to cartilage damage in their right knee and had a femoral shaft osteotomy to remove pressure from the damaged tissue. The fixation metallic implant screwed to the femur is cleanly visible in a lateral X-ray computed radiographic image (Fig.\,\ref{fig:Implants}(c)), but leads to high intensity fringes around the metallic hardware in high-field MR images due to incorrect spin mapping (see Fig.\,\ref{fig:Implants}(d), taken at 3\,T). These effects depend supralinearly on the magnetic field strength and are barely perceptible\cite{VanSpeybroeck2021} at fields $<0.1$\,T. The field dependence is notorious in the images: the SNR and resolution are much higher in the 3\,T system, but the metallic implant geometry is accurately defined in our 70\,mT 2D and 3D reconstructions, and can be readily segmented with standard data post-processing. The low-field images were taken in 12\,min (Fig.\,\ref{fig:Implants}(a) and (b)) and 20\,min (Fig.\,\ref{fig:Implants}(e)) with $T_1$-weighted 3D-TSE acquisitions (Methods).

\begin{figure}
	\centering
	\includegraphics[width=1.\columnwidth]{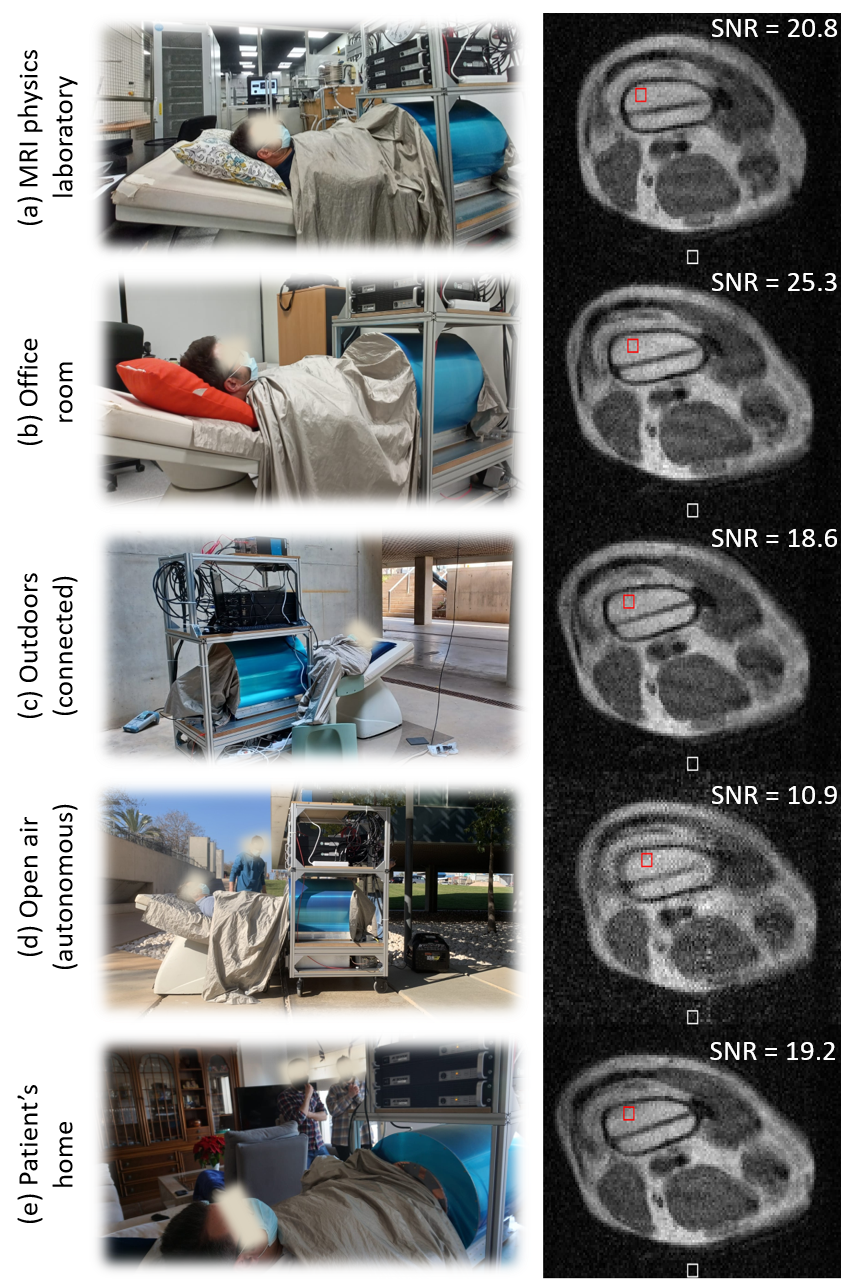}
	\caption{Photographs during acquisitions (left) and axial slice from 3D-TSE reconstructions (right, no post-processing) at five different locations: (a) in an MRI physics laboratory; (b) in an office room; (c) outside a campus building, connected to a nearby power outlet; (d) in open air, powered from a small fuel-based generator; and (e) at the volunteer's home.}
	\label{fig:Locations}
\end{figure}

The goal of the last set of experiments is to evaluate the portability of the scanner and its performance under various environments and conditions. To establish comparisons as unbiased as possible, the acquisitions across all five scenarios are of the intervened knee of the same volunteer as in Fig.\,\ref{fig:Implants}, and all with the same sequence parameters: $T_1$-weighted 3D-TSE with a total scan time $\approx12$\,min (Methods). The slices in Fig.\,\ref{fig:Locations} have been selected to show the third screw from the top (Fig.\,\ref{fig:Implants}) as it runs through the bone from the top of the image, where the prosthetic plate is implanted. As a general indicator of the image quality, we measure the SNR in a region of interest (ROI) in the femur bone marrow (red boxes in the unprocessed reconstructions in Fig.\,\ref{fig:Locations}). To this end, we estimate the signal strength as the average voxel brightness in the ROI, and the noise as the average voxel brightness in the background (white boxes). Prior to each acquisition, we measured the spectral noise density as picked up by the detection RF coil with the subject inside the scanner. The average signal strength in these spectra speaks of the white noise amplitude in the Rx chain, which is ideally close to thermal (Johnson) noise in the coil (Methods). Besides, we often encounter stronger peaks, indicating EMI at discrete frequencies. These can be suppressed by covering the subject meticulously with the shielding cloth.

The first acquisition (Fig.\,\ref{fig:Locations}(a)) took place in the same laboratory as above and serves as reference under controlled ambient conditions. For the acquisition in the MRI physics laboratory, the Larmor frequency was $\approx3.076$\,MHz, the measured noise level (50\,nV/Hz$^{1/2}$) was compatible with Johnson noise (Methods), there is no visible EMI and the femur SNR is $\approx 21$. 

The second scan took place in an office room (Fig.\,\ref{fig:Locations}(b)) around 20\,m away from the laboratory, in the same building and floor. The Larmor frequency here decreased to $\approx 3.064$\,MHz due to a higher temperature. The noise amplitude is still consistent with Johnson noise levels and EMI is not visible in the reconstruction. The SNR in the marrow ROI is $\approx 25$ and the overall image quality is comparable to the reference image, perhaps even slightly sharper.

The third image was acquired outdoors, at basement level, just outside the laboratory building (Fig.\,\ref{fig:Locations}(c)). The system was powered through a 30\,m cable running down three floors from the laboratory. The conducting cloth wrapped around the subject purposely connected the scanner shielding to the concrete floor to improve the otherwise resistive connection between the laboratory ground and earth. During this acquisition the volunteer reported sensing the presence and conversations of bypassers, a light breeze on the grounded cloth, and weak tremors due to vehicles driving through the underground parking. The resulting image quality seems not to be strongly influenced by any of these, with an SNR of $\approx 19$ in the ROI, and a noise spectrum of comparable amplitude to indoor acquisitions. The Larmor frequency was $\approx 3.065$\,MHz.

The fourth scan was also taken outdoors, in this case in open air in a university campus (Fig.\,\ref{fig:Locations}(d), Larmor frequency $\approx 3.063$\,MHz), far from power outlets and operating autonomously with a portable electricity generator. The latter is based on a low-consumption gasoline engine, weighs $<20$\,kg, costs $<600$\,\euro{} and has an autonomy $>10$\,hours with the scanner at continuous operation (Methods). We grounded the system electrically as before, with the conducting cloth offering low-resistance paths between the scanner shielding, the floor concrete and the ground terminal in the generator. The spectrum was significantly more populated in this case, presumably due to noise originating at the motor. Consequently, the quality of the resulting image is somewhat lower than in the previous acquisitions ($\text{SNR}\approx 11$), and an EMI line is visible along the vertical (phase-encoding) direction. Nevertheless, the main anatomic features, different tissues and metallic implants are all still clearly identifiable.

The last image was acquired at the volunteer's apartment. This is located in a low-density town in the province of Valencia, Spain. The system was transported in a small truck from the university campus to a parking lot approximately 300\,m away from the entrance to the apartment block, and pushed along the sidewalk into the building, the elevator, the apartment and ultimately, the living room. Throughout the way, the only wheelchair-adapted elements were the lowered-sidewalks at pedestrian crossings. After transport, the system required re-connecting some of the RF electronics modules we had packed in a separate box, and tightening some screwed connectors that had become loose during transport through the rugged, tiled sidewalks. Other than that, the system was plugged to a wall power outlet, tuned to $\approx 3.065$\,MHz and ready to use. The noise spectrum in the apartment was clean, again compatible with Johnson noise levels. The SNR in the ROI for this acquisition is $\approx 19$.

In conclusion, we have demonstrated the viability of a portable, low-cost system for magnetic resonance imaging indoors, outdoors and at home. In this work, we have focused on healthy volunteers and subjects carrying metallic implants. Nevertheless, the acquired images contain sufficient anatomical information to diagnose a large variety of articular diseases, including effusion, synovial engorgement, tendon disruption or bone fractures. Upon completion of this manuscript, we have learned about an independent effort towards portable MRI with a commercial 64\,mT scanner mounted on a modified cargo van\cite{Deoni2021}.

Looking ahead, our 70\,mT scanner can be still upgraded in various ways. Machine learning algorithms have been shown to boost the performance in other low-field systems and can be readily incorporated to ours. These can be used, via transfer learning, to increase the spatial resolution of scans \emph{a posteriori} based on multiple acquisitions, prior knowledge about the sample\cite{VanReeth2012}, or with networks trained with paired datasets of low and high-field images, to recover from the former features visible otherwise only with the latter\cite{GarciaHernandez2021,Iglesias2022}. Deep learning and convolutional neural networks can also be employed to increase reconstruction quality through image denoising, artifact detection and active noise cancellation\cite{Koonjoo2021,Kustner2018,Liu2021}. Quantitative MRI, radiomics and fingerprinting\cite{OReilly2021,Sarracanie2021,Simpson2020} show promising potential in situations where subtle texture changes contain potentially valuable information for the patients. Also special-purpose pulse sequences and reconstruction methods can enhance the efficiency of low-field MRI\cite{Sarracanie2021,Algarin2020}, and hardware developments and contrast agents which are mainstays in clinical high-field MRI (e.g. parallel imaging, optimization of RF detection coils for different body parts, gadolinium contrast enhancement), are seldom used in the still mostly experimental low-field systems available\cite{Marques2019,Waddington2020}. Finally, for our particular scanner, the GUI and overall system usability can be improved to facilitate operation by non-experts.

All in all, the scanner performance demonstrated in this work, especially if upgraded with the above capabilities, sets a path towards accessible MRI, democratizing its use and benefits, and qualitatively expanding the circumstances where it can provide value.

\backmatter

%

\bmhead{Acknowledgments}

We thank the anonymous volunteers for their participation, Andrew Webb and Thomas O'Reilly for discussions on MRI hardware, and Benjamin Menk\"uc for contributions to MaRCoS.

\bmhead{Funding}

This work was supported by the Ministerio de Ciencia e Innovaci\'on of Spain through research grant PID2019-111436RB-C21. Action co-financed by the European Union through the Programa Operativo del Fondo Europeo de Desarrollo Regional (FEDER) of the Comunitat Valenciana 2014-2020 (IDIFEDER/2018/022 and IDIFEDER/2021/004) and Future and Emerging Technologies (FET, grant 101034644). JMG and JB acknowledge support from the Innodocto program of the Agencia Valenciana de la Innovaci\'on (INNTA3/2020/22 and INNTA3/2021/17).

\bmhead{Conflict of interest}

PhysioMRI Tech S.L. is a for-profit organization spun off the Institute for Molecular Imaging and Instrumentation and proprietor of the low-field scanner presented in this work. JMA, FG, JB, JMB and JA have patents pending that are licensed to PhysioMRI Tech S.L. JMA, FG, AR, JMB and JA are co-founders of PhysioMRI Tech S.L. All other authors declare no competing interests.

\bmhead{Ethics approval}

All experiments were carried out following Spanish regulations and under the research agreement from La Fe Hospital in Valencia (IIS-F-PG-22-02, agreement number 2019-139-1).

\bmhead{Consent to participate}

Informed consent was obtained from all volunteers prior to study commencement.

\bmhead{Consent for publication}

Informed consent for publication was obtained from all participants prior to study commencement. 

\bmhead{Availability of data and materials}

All anonymized datasets, reconstruction and postprocessing methods generated and/or used during the present study are available from the corresponding author upon reasonable request.

\bmhead{Code availability}

MaRCoS and the GUI are publicly available from open-source repositories at \url{https://github.com/vnegnev/marcos_server} and \url{https://github.com/yvives/PhysioMRI_GUI}, respectively.

\bmhead{Authors' contributions}

Low-field images taken by TGN and JMA with help from RB, PM, FJL, JPR and JA. Portable system built by RPG, TGN, JMA, FG, RB, EP, JMG and JA with help from JPR, PM and FJL. Data analysis and evaluation by JMA, TGN, FG, LMB and JA. Control electronics and software developed by VN, YVG, JMA, TGN and JB. Portability experiments conceived by JA, JMA, TGN and AMP. Project conceived and supervised by AR, JMB and JA. Paper written by JA and LMB, with input from all authors.

\begin{appendices}

%




\end{appendices}




\section*{Methods}
\subsection*{Scanner}
The scanner is based on a Halbach magnet including almost 4,600 NdFeB cubes of side 12\,mm to generate $B_0\approx 72$\,mT at the field of view, and another $\approx1,100$ smaller cuboids (64\,mm$^3$) to shim the inhomogeneity from $\approx15,700$ down to $\approx3,100$\,ppm over a spherical volume of 20\,cm in diameter. The gradient coil geometry is optimized with conventional target-field methods for efficiencies between 0.52 and 0.91\,mT/m/A. These coils are wound on and glued to curved 3D-printed Nylon molds, and the whole assembly is supported by a methacrylate cylinder. Gradient waveforms are generated with an OCRA1 board connected to the Red Pitaya in the MaRCoS system via Serial Peripheral Interface (SPI), and amplified by AE Techron 7224 power amplifiers (Indiana, USA). The single Tx/Rx RF antenna is a solenoid coil tuned and impedance-matched to the proton Larmor frequency ($\approx3$\,MHz). The RF coil holder was 3D-printed in polylactic acid (PLA), and the wire was fixed with cyanoacrylate adhesive. The coil is inside a grounded faraday cage for noise immunity and to prevent interference between the gradients and the RF system, and a conductive cloth covers the subject during \emph{in vivo} acquisitions. The RF low-noise and power amplifiers, as well as the Tx/Rx switch, were purchased from Barthel HF-Technik GmbH (Aachen, Germany).

\subsection*{Pulse sequences}
The knee image in Fig.\,\ref{fig:KneeHand}(a) was acquired with a $T_1$-weighted 3D-TSE sequence, with $\text{FoV} = 130\times 140\times 180$\,mm$^3$, a pixel resolution of $1.85\times 1.75\times 2$\,mm$^3$, $\text{ETL} = 5$, $\text{TE} = 20$\,ms, $\text{TR} = 200$\,ms, $\text{BW} = 17.5$\,kHz, and 4 averages for a total scan time of 19.2\,min. The duration of resonant $\pi/2$ and $\pi$-pulses in all images are $\approx\SI{40}{\micro s}$ and $\approx\SI{80}{\micro s}$, respectively, and dephasing gradient pulses after the RF pulses are pre-emphasized by a factor $\approx 1.006$ to place the echoes at the center of the data acquisition windows.

The hand image in Fig.\,\ref{fig:KneeHand}(b) was acquired with a $T_1$-weighted 3D-TSE sequence, with $\text{FoV} = 180\times 180\times 50$\,mm$^3$, a pixel resolution of $1.5\times 1.5\times 5$\,mm$^3$, $\text{ETL} = 10$, $\text{TE} = 20$\,ms, $\text{TR} = 400$\,ms, $\text{BW} = 30$\,kHz, and 13 averages for a total scan time of 10.4\,min.

The wrist image in Fig.\,\ref{fig:KneeHand}(c) was acquired with a $T_1$-weighted 3D-TSE sequence, with $\text{FoV} = 180\times 140\times 80$\,mm$^3$, a pixel resolution of $1.5\times 1.5\times 10$\,mm$^3$, $\text{ETL} = 3$, $\text{TE} = 20$\,ms, $\text{TR} = 100$\,ms, $\text{BW} = 30$\,kHz, and 30 averages for a total scan time of 12\,min. 

The wrist image in Fig.\,\ref{fig:KneeHand}(d) was acquired with a $\rho$-weighted 3D-TSE sequence, with $\text{FoV} = 180\times 140\times 80$\,mm$^3$, a pixel resolution of $1.5\times 1.5\times 10$\,mm$^3$, $\text{ETL} = 5$, $\text{TE} = 20$\,ms, $\text{TR} = 1000$\,ms, $\text{BW} = 30$\,kHz, and 5 averages for a total scan time of 12\,min. 

The wrist image in Fig.\,\ref{fig:KneeHand}(e) was acquired with a $T_2$-weighted 3D-TSE sequence, with $\text{FoV} = 180\times 140\times 80$\,mm$^3$, a pixel resolution of $1.5\times 1.5\times 10$\,mm$^3$, $\text{ETL} = 5$, echo spacing of 20\,ms, effective $\text{TE} = 100$\,ms, $\text{TR} = 1000$\,ms, $\text{BW} = 30$\,kHz, and 5 averages for a total scan time of 12\,min. 

The knee images in Fig.\,\ref{fig:Implants}(a) and (b) were acquired with a $T_1$-weighted 3D-TSE sequence, with $\text{FoV} = 200\times 200\times 180$\,mm$^3$, a pixel resolution of $1.3\times 2\times 9$\,mm$^3$, $\text{ETL} = 5$, $\text{TE} = 20$\,ms, $\text{TR} = 200$\,ms, $\text{BW} = 37.5$\,kHz, and 9 averages for a total scan time of 12\,min.

The knee image in Fig.\,\ref{fig:Implants}(e) was acquired with a $T_1$-weighted 3D-TSE sequence, with $\text{FoV} = 200\times 200\times 180$\,mm$^3$, a pixel resolution of $2\times 2\times 2$\,mm$^3$, $\text{ETL} = 10$, $\text{TE} = 20$\,ms, $\text{TR} = 300$\,ms, $\text{BW} = 22.5$\,kHz, and 4 averages for a total scan time of 20\,min.

The knee images in Fig.\,\ref{fig:Locations} were acquired with a $T_1$-weighted 3D-TSE sequence, with $\text{FoV} = 180\times 200\times 200$\,mm$^3$, a pixel resolution of $1.2\times 2\times 10$\,mm$^3$, $\text{ETL} = 5$, $\text{TE} = 20$\,ms, $\text{TR} = 200$\,ms, $\text{BW} = 37.5$\,kHz, and 9 averages for a total scan time of 12\,min.

\subsection*{Data acquisition}
The receive chain consists of an analog stage (RF coil, passive Tx/Rx switch, low-noise amplifier and low-pass filter) followed by a digital stage. The digitization is performed at 122.88\,Ms/s by an analog-to-digital converter in the Red Pitaya Stemlab board of MaRCoS. The digital signal is mixed down by complex multiplication with a numerically-controlled oscillator set to the Larmor frequency. The real and imaginary data components pass first a cascaded integrator-comb filter and finally a finite impulse response filter. The resulting data conform the sought in-phase and quadrature components of the magnetic resonance signal. These are sent to the control computer and can be Fourier-transformed for image reconstruction and post-processing.

\subsection*{Noise}
The spectral noise density of the MR data is bounded from below by Johnson noise due to thermal fluctuations of electrons in the resistive elements $R$ in the detector. These are dominated by the coil, with quality factor $Q\approx 93$ (88) and $R\approx 5$ $(\SI{5.5}{\ohm})$ in the unloaded (loaded) case. For a given acquisition bandwidth, the integrated noise amplitude is expected to be $(4k_\text{B} R\cdot BW)^{1/2}$, with $k_\text{B}$ the Boltzmann constant. In the controlled environment of the MRI physics laboratory, we measure $\approx 50$\,nV/Hz$^{1/2}$ after a 45\,dB low-noise pre-amplifier, in agreement with the estimated Johnson level. We use this as a reference to evaluate the signal quality and the shielding efficiency of the conductive cloth, both in the laboratory and in the rest of locations.

We have found situations where suppressing noise down to Johnson levels is not trivial, and indeed did not achieve it when the system was powered by the portable generator. The control computer is another significant source of 50\,Hz noise and needs to be as far as possible in the rack to reconstruct clean images. We also find it often necessary to ensure the subject is sufficiently covered by the conductive cloth, and extending some of it on the floor helps. 

\subsection*{Generator}
For the autonomous experiments outdoors we powered the system from a ``Limited 2000i'' gasoline-fueled generator from Genergy (Calahorra, Spain). This motor delivers up to 2\,kW at 230\,V and 50\,Hz (single phase). It costs $<600$\,\euro, weighs 19\,kg and has a fuel tank capacity of 4\,l and an autonomy of 10.8 hours at 25\,\% load (500\,W), which is more than required for continuous operation of the scanner.

\end{document}